\documentclass[twocolumn,aps,superscriptaddress,showpacs]{revtex4}
\usepackage{amssymb}
\usepackage{amsmath}
\usepackage{graphicx}
\usepackage[normalem]{ulem}
\usepackage[dvips]{color}

\begin{document}

\title{Specific viscosity of neutron-rich nuclear matter from the relaxation time approach}
\author{Jun Xu}\email{xujun@comp.tamu.edu}
\affiliation{Cyclotron Institute, Texas A\&M University, College
Station, Texas 77843-3366, USA}

\date{\today}

\begin{abstract}

The specific viscosity of neutron-rich nuclear matter is studied
from the relaxation time approach using an isospin- and
momentum-dependent interaction and the nucleon-nucleon cross
sections taken as those from the experimental data modified by the
in-medium effective masses as used in the IBUU transport model
calculations. The relaxation time of neutrons is larger while that
of protons is smaller in neutron-rich nuclear matter compared with
that in symmetric nuclear matter, and this leads to a larger
specific viscosity in neutron-rich nuclear matter. In addition, the
specific viscosity decreases with increasing temperature because of
more frequent collisions and weaker Pauli blocking effect at higher
temperatures. At lower temperatures the specific viscosity increases
with increasing density due to the Pauli blocking effect, while at
higher temperatures it slightly decreases with increasing density as
a result of smaller in-medium effective masses at higher densities.

\end{abstract}


\maketitle

\section{Introduction}
\label{introduction}

The properties of nuclear matter under extreme conditions is one of
the major problems in nuclear physics. Transport models are now
widely used in studying the equation of states (EOS) of the nuclear
matter formed in intermediate-energy heavy-ion
collisions~\cite{Ber88}. It was found that the transverse flow from
the experimental data can be reproduced by the transport model
calculation using a stiff EOS together with a momentum-independent
potential or a soft EOS together with a momentum-dependent
potential~\cite{GBD87}. However, the extracted incompressibility
from Giant Monopole Resonance experiments is $231\pm5$
MeV~\cite{You99}, indicating a soft EOS and the importance of the
momentum-dependent mean field potential. Recently, an isospin- and
momentum-dependent interaction together with the Isospin-dependent
Boltzmann-Uehling-Uhlenbeck (IBUU) transport model has been
extensively used to study the dynamics in intermediate-energy
heavy-ion collisions. Especially, extensive studies have been done
to extract the information on the isospin-dependent part of the EOS,
i.e., symmetry energy, from isospin diffusion~\cite{Che05},
neutron/proton~\cite{Yon06} and triton/$^3$He~\cite{Yon09} ratios
and differential flows, and $\pi^-/\pi^+$ ratio~\cite{Xia09}. For a
recent review, I refer the readers to Ref.~\cite{Li08}.

Although hydrodynamic models have seldom been used in
intermediate-energy heavy-ion collisions~\cite{Sch93}, they are
widely used in relativistic heavy-ion collisions, where a very
strong interacting matter called quark-gluon plasma (QGP) is formed.
It has been found that this matter has a very small viscosity and
behaves like a nearly perfect fluid. Below the temperature
$T_c=170\sim180$ MeV hadronization happens and the viscosity of the
system increases. Although there have already been extensive studies
on the shear viscosity of the QGP~\cite{Pes05,Maj07,Xu08,Che10} and
that of the relativistic hadron
gas~\cite{Mur04,Che07a,Dem09,Pal10a}, only a few studies have been
devoted to the viscosity of nuclear matter formed in
intermediate-energy heavy-ion
collisions~\cite{Dan84,Shi03,Che07b,Pal10b,Li11}. Even fewer studies
have been devoted to the isospin dependence of the viscosity
~\cite{Zha10}. It is thus of great interesting to study the
viscosity of the neutron-rich nuclear matter, especially its isospin
dependence, formed in the intermediate-energy heavy-ion collisions
as in the IBUU transport model.

Many methods in the literature have been used to study the
viscosity~\cite{Dan84,Jeo95,Che07a,Xu08}, among which directly using
Green-Kubo formulas is the first-principle way of the
study~\cite{Kub66}. In the present study the relaxation time
approach is used~\cite{Hua87,Abu93}. This approach is helpful in
studying how the Pauli blocking, in-medium cross sections and
in-medium effective masses will affect the shear viscosity and can
give qualitative ideas how the shear viscosity changes with density,
temperature and isospin asymmetry of the nuclear matter.

This paper is organized as follows. The isospin- and
momentum-dependent interaction (here after 'MDI') is briefly
reviewed in Sec.~\ref{MDI}. The method to calculate the shear
viscosity from the relaxation time approach is given in
Sec.~\ref{viscosity}. In Sec.~\ref{results} the results of the
relaxation time and the specific viscosity are shown and the effects
from isospin, temperature and Pauli blocking are discussed. A
summary is given in Sec.\ref{summary}.

\section{The isospin- and momentum-dependent interaction}
\label{MDI}

The MDI interaction is a modified finite-range Gogny-like
interaction~\cite{Das03}, and it was recently found that the
effective nucleon-nucleon potential is composed of a zero-range
many-body term and a finite-range Yukawa
term~\cite{Xu10}. In the mean-field approximation, the single-particle potential for a nucleon with momentum $\vec{p}$ and isospin $%
\tau $ in a nuclear matter medium is written as
\begin{eqnarray}
U_\tau(\vec{p}) &=&A_{u}(x)\frac{\rho _{-\tau }}{\rho _{0}}%
+A_{l}(x)\frac{\rho _{\tau }}{\rho _{0}}  \notag \\
&+&B(\frac{\rho }{\rho _{0}})^{\sigma }(1-x\delta ^{2})-8\tau x\frac{B}{%
\sigma +1}\frac{\rho ^{\sigma -1}}{\rho _{0}^{\sigma }}\delta \rho _{-\tau }
\notag \\
&+&\frac{2C_{\tau ,\tau }}{\rho _{0}}\int \frac{d^{3}p^{\prime
}}{(2\pi)^3} \frac{f_{\tau } (\vec{p}^{\prime
})}{1+(\vec{p}-\vec{p}^{\prime })^{2}/\Lambda ^{2}}
\notag \\
&+&\frac{2C_{\tau ,-\tau }}{\rho _{0}}\int
\frac{d^{3}p^{\prime}}{(2\pi)^3}\frac{f_{-\tau } (\vec{p}^{\prime
})}{1+(\vec{p}-\vec{p}^{\prime })^{2}/\Lambda ^{2}}. \label{MDIU}
\end{eqnarray}%
In the above $\tau =1/2$ ($-1/2$) for neutrons (protons);
$\delta=(\rho_n-\rho_p)/\rho$ is the isospin asymmetry with
$\rho_n$($\rho_p$) being the neutron (proton) density and
$\rho=\rho_n+\rho_p$ being the total density of the medium; $f_{\tau
}(\vec{p})$ is the local phase space distribution function and it
can be written as
\begin{equation}\label{f}
f_{\tau }(\vec{p})=dn_\tau(\vec{p}) = \frac{d}{\exp \left[(\frac{p^{2}}{%
2m}+U_{\tau }(\vec{p})-\mu_\tau )/T\right]+1},
\end{equation}
where $d=2$ is the spin degeneracy and $n_\tau(\vec{p})$ is the
occupation probability. $m$ is the nucleon mass, $T$ is the
temperature and $\mu _{\tau }$ is the proton or neutron chemical
potential and can be
determined from%
\begin{equation}
\rho _{\tau }=\int f_{\tau }(\vec{p})\frac{d^{3}p}{(2\pi)^3}.
\end{equation}%
From a self-consistent iteration method~\cite{Xu07}, both the
single-particle potential and the distribution function can be
calculated numerically. The entropy density of the nuclear matter
can thus be calculated from
\begin{equation}
s =-\sum_\tau d\int [n_{\tau }\ln n_{\tau }+(1-n_{\tau })\ln
(1-n_{\tau })]\frac{d^3p}{(2\pi)^3} \label{S}
\end{equation}%

The detailed values of the parameters $A_{u}(x)$, $A_{l}(x)$,
$\sigma$, $B$, $C_{\tau ,\tau }$, $C_{\tau ,-\tau }$ and $\Lambda $
can be found in Ref.~\cite{Das03} and they are assumed to be
independent of the temperature. These parameters lead to the binding
energy $E_0=-16$ MeV, the incompressibility $K_0=211$ and the
symmetry energy $E_{sym}\approx31$ MeV at the saturation density
$\rho_0=0.16$ fm$^{-3}$. The parameter $x$ is used to mimic the
density dependence of the symmetry energy while keeping the
properties of symmetric nuclear matter unchanged. Comparison between
the results from the IBUU transport model calculation with the
experimental results of isospin diffusion data leads to $-1<x<0$ at
subsaturation densities~\cite{Che05,Li05} and with that of
$\pi^-/\pi^+$ ratio somehow leads to a super soft symmetry energy
$x=1$ at suprasaturation densities~\cite{Xia09}. In the following
$x=0$ is used if not addressed.

\begin{figure}[tbh]
\includegraphics[scale=0.8]{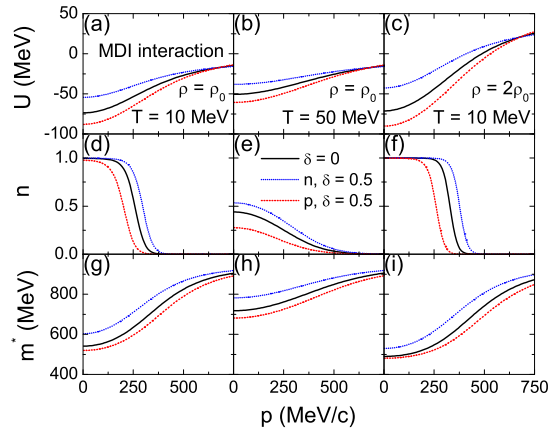}
\caption{(color online) Momentum dependence of the single-particle
potential, the occupation probability and the effective mass at
$\rho=\rho_0$ and $T=10$ MeV ((a), (d), (g)), $\rho=\rho_0$ and
$T=50$ MeV ((b), (e), (h)), and $\rho=2\rho_0$ and $T=10$ MeV ((c),
(f), (i)) from the MDI interaction. Results from symmetric
($\delta=0$) nuclear matter and neutron-rich ($\delta=0.5$) nuclear
matter are compared.} \label{spp}
\end{figure}

Once the momentum dependence of the single-particle potential is obtained, the effective mass of a nucleon
with isospin $\tau$ in the nuclear matter medium can be calculated from
\begin{equation}\label{effm}
\frac{1}{m_\tau^\star} = \frac{1}{m} + \frac{1}{p}
\frac{dU_\tau}{dp}.
\end{equation}
In Fig.~\ref{spp} the momentum dependence of the single-particle
potential, the occupation probability and the effective mass are
displayed at different densities and temperatures. It is seen that
the momentum-dependence of the single-particle potential is weaker
at higher temperatures, which leads to a larger effective mass with
increasing temperature. A higher temperature also leads to a more
diffusive Fermi surface. At higher densities, the effective mass is
smaller and the distribution function becomes less diffusive. The
effective mass of neutrons is larger than that of protons in
neutron-rich nuclear matter from the MDI interaction. This leads to
a symmetry potential $U_{sym}=(U_n-U_p)/\delta$ that is consistent
with the energy dependence of the Lane potential constrained by the
nucleon-nucleus scattering experimental data as discussed in
Ref.~\cite{Li04}. The relation between the shear viscosity and these
single-particle properties will be discussed in the following.

\section{Shear viscosity from the relaxation time approach}
\label{viscosity}

In the present work I study the shear viscosity of an isospin
asymmetric nuclear matter with uniform and static neutron and proton
density $\rho_n$ and $\rho_p$, respectively, and temperature $T$.
The static flow field in the nuclear system is assumed to be in the
$z$ direction and the magnitude changes linearly with $x$, i.e.,
$u_z=cx$, $u_x=u_y=0$. The equilibrium momentum distribution $n^0$
is a Fermi-Dirac distribution as Eq.~(\ref{f}) in the local frame
while it is shifted by $m\vec{u}$ with the flow field in the lab
frame. Due to collisions, the momentum distribution may differ
slightly from the equilibrium one and is written as $n$ and $\delta
n = n - n^0$ is their difference.

The shear force between flow layers per unit area is~\cite{Hua87}
\begin{eqnarray}
F &=& \sum_\tau <(p_z-mu_z)\rho_\tau \frac{p_x}{m_\tau^\star}>
\notag\\
&=& \sum_\tau d \int (p_z-mu_z)\frac{p_x}{m_\tau^\star}n_\tau
\frac{d^3p}{(2\pi)^3}.
\end{eqnarray}
In the above $\rho_\tau\frac{p_x}{m_\tau^\star}$ is the flux, i.e.,
the number of nucleons of isospin $\tau$ moving in the x direction
per unit time per unit area, where $\frac{p_x}{m_\tau^\star}$ can be
written as $\left[\frac{p_x}{m} + (\nabla_p U_\tau)_x\right]$ from
Eq.~(\ref{effm}), and $p_z-mu_z$ is the momentum transported per
nucleon. It is easily seen that the contribution of $n^0_\tau$ to
the shear force is zero as the integrand is odd in $p_x$. Thus, the
shear force can be written as
\begin{equation}\label{shearF}
F = \sum_\tau \frac{d}{(2\pi)^3} \int
(p_z-mu_z)\frac{p_x}{m^\star_\tau} \delta n_\tau dp_x dp_y dp_z.
\end{equation}
\begin{widetext}
To calculate $\delta n_\tau$, we start from the BUU equation using
the relaxation time approximation. The isospin-dependent BUU
equation is written as
\begin{eqnarray}
&&\frac{\partial n_\tau(p_1)}{\partial t} + \vec{v} \cdot \nabla_r
n_\tau(p_1) - \nabla_r U_\tau \cdot \nabla_p n_\tau(p_1) = -
(d-\frac{1}{2})\int \frac{d^3p_2}{(2\pi)^3}
\frac{d^3p_1^\prime}{(2\pi)^3} \frac{d^3p_2^\prime}{(2\pi)^3}
|T_{\tau,\tau}|^2 \notag \\
&\times&
[n_\tau(p_1)n_\tau(p_2)(1-n_\tau(p_1^\prime))(1-n_\tau(p_2^\prime))-
n_\tau(p_1^\prime)n_\tau(p_2^\prime)(1-n_\tau(p_1))(1-n_\tau(p_2))] \notag\\
&\times& (2\pi)^3
\delta^{(3)}(\vec{p}_1+\vec{p}_2-\vec{p}_1^\prime-\vec{p}_2^\prime)-
d\int \frac{d^3p_2}{(2\pi)^3} \frac{d^3p_1^\prime}{(2\pi)^3}
\frac{d^3p_2^\prime}{(2\pi)^3}
|T_{\tau,-\tau}|^2 \notag \\
&\times&
[n_\tau(p_1)n_{-\tau}(p_2)(1-n_\tau(p_1^\prime))(1-n_{-\tau}(p_2^\prime))-
n_\tau(p_1^\prime)n_{-\tau}(p_2^\prime)(1-n_\tau(p_1))(1-n_{-\tau}(p_2))] \notag\\
&\times& (2\pi)^3
\delta^{(3)}(\vec{p}_1+\vec{p}_2-\vec{p}_1^\prime-\vec{p}_2^\prime),
\end{eqnarray}
\end{widetext}
where the terms $1-n$ are from the Pauli blocking effect and the
degeneracy $d-\frac{1}{2}$ takes identical nucleon collisions into
account. In the first-order approximation, the contribution of
$\delta n_\tau$ on the left side can be neglected and $n_\tau$ is
replaced by $n^0_\tau$. The first term on the left side vanishes as
$n^0_\tau$ is the equilibrium distribution and the system is static.
By using the relation
\begin{eqnarray}
\nabla_r n^0_\tau &=& \frac{\partial n^0_\tau}{\partial x} \hat{x} \notag\\
&=& \frac{n^0_\tau(p_x,p_y,p_z-mc\delta x) -
n^0_\tau(p_x,p_y,p_z)}{\delta x} \hat{x}
\notag\\
&=& -mc\frac{p_z}{p}\frac{dn^0_\tau}{dp}\hat{x},
\end{eqnarray}
the second term can be calculated as
\begin{equation}
\vec{v} \cdot \nabla_r n^0_\tau = \left(\frac{p_x}{m} +
\frac{dU^0_\tau}{dp}\frac{p_x}{p}\right)
\left(-mc\frac{p_z}{p}\frac{dn^0_\tau}{dp}\right),
\end{equation}
where $U^0_\tau$ is the single-particle potential corresponding to
the equilibrium distribution $n^0_\tau$. The third term on the left
side can be similarly expressed as
\begin{eqnarray}
-\nabla_r U^0_\tau \cdot \nabla_p n^0_\tau =
-\left(-mc\frac{p_z}{p}\frac{dU^0_\tau}{dp}\right)
\left(\frac{p_x}{p}\frac{dn^0_\tau}{dp}\right).
\end{eqnarray}
So, the left side can be expressed as
\begin{eqnarray}\label{LHS}
&&\frac{\partial n_\tau(p_1)}{\partial t} + \vec{v} \cdot \nabla_r
n_\tau(p_1) - \nabla_r U_\tau \cdot \nabla_p n_\tau(p_1) \notag\\
&=& \left(-\frac{\partial u_z}{\partial x} \frac{p_z p_x}{p} \frac{d
n^0_\tau}{dp}\right)_{p=p_1},
\end{eqnarray}
as $c=\partial u_z/\partial x$.

\begin{figure}[tbh]
\includegraphics[scale=0.8]{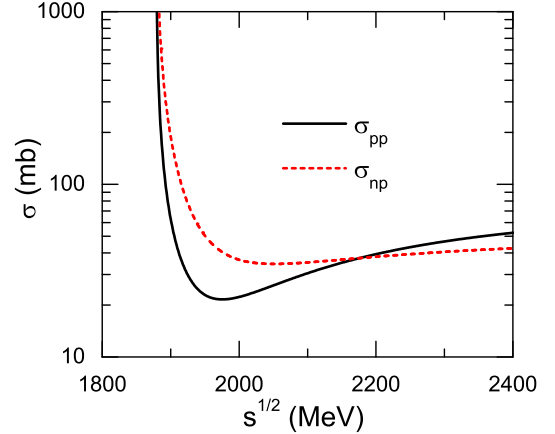}
\caption{(color online) $pp$ and $np$ cross sections in free space
as functions of the center-of-mass energy of the two colliding
nucleons from the parameterized form (Eqs.~(\ref{sigma1}) and
(\ref{sigma2})).} \label{freecs}
\end{figure}

\begin{widetext}
The right side vanishes if $n_\tau=n^0_\tau$ when the detailed
balance is satisfied. In the relaxation time approximation only
$\delta n_\tau(p_1)$ is kept and the right side can be written
as~\cite{Hua87}
\begin{eqnarray}\label{RHS}
-\frac{\delta n_\tau(p_1)}{\tau_\tau(p_1)}&=&- (d-\frac{1}{2})\int
\frac{d^3p_2}{(2\pi)^3} \frac{d^3p_1^\prime}{(2\pi)^3}
\frac{d^3p_2^\prime}{(2\pi)^3}
|T_{\tau,\tau}|^2 \notag \\
&\times& [\delta
n_\tau(p_1)n^0_\tau(p_2)(1-n^0_\tau(p_1^\prime))(1-n^0_\tau(p_2^\prime))+
n^0_\tau(p_1^\prime)n^0_\tau(p_2^\prime)\delta n_\tau(p_1)(1-n^0_\tau(p_2))] \notag\\
&\times& (2\pi)^3
\delta^{(3)}(\vec{p}_1+\vec{p}_2-\vec{p}_1^\prime-\vec{p}_2^\prime)-
d\int \frac{d^3p_2}{(2\pi)^3} \frac{d^3p_1^\prime}{(2\pi)^3}
\frac{d^3p_2^\prime}{(2\pi)^3}
|T_{\tau,-\tau}|^2 \notag \\
&\times& [\delta
n_\tau(p_1)n^0_{-\tau}(p_2)(1-n^0_\tau(p_1^\prime))(1-n^0_{-\tau}(p_2^\prime))+
n^0_\tau(p_1^\prime)n^0_{-\tau}(p_2^\prime)\delta n_\tau(p_1)(1-n^0_{-\tau}(p_2))] \notag\\
&\times& (2\pi)^3
\delta^{(3)}(\vec{p}_1+\vec{p}_2-\vec{p}_1^\prime-\vec{p}_2^\prime),
\end{eqnarray}
where $\tau_\tau(p_1)$ is the relaxation time, i.e., the average
collision time for a nucleon with isospin $\tau$ and momentum $p_1$,
which can be expressed as
\begin{equation}
\frac{1}{\tau_\tau(p_1)} = \frac{1}{\tau_\tau^{same}(p_1)} +
\frac{1}{\tau_\tau^{diff}(p_1)},
\end{equation}
with $\tau_\tau^{same(diff)}(p_1)$ being the average collision time
for a nucleon with isospin $\tau$ and momentum $p_1$ when colliding
with other nucleons of same (different) isospin, and they can thus
be calculated from
\begin{eqnarray}
\frac{1}{\tau_\tau^{same}(p_1)} &=&
(d-\frac{1}{2})\frac{(2\pi)^2}{(2\pi)^3} \int p_2^2 dp_2
d\cos\theta_{12} d\cos\theta
\frac{d\sigma_{\tau,\tau}}{d\Omega}\left|\frac{\vec{p}_1}{m_\tau^\star(p_1)}-\frac{\vec{p}_2}{m_\tau^\star(p_2)}\right|
\notag\\
&\times&\left[n^0_{\tau}(p_2)-n^0_{\tau}(p_2)n^0_\tau(p_1^\prime)-n^0_{\tau}(p_2)n^0_{\tau}(p_2^\prime)+n^0_\tau(p_1^\prime)n^0_{\tau}(p_2^\prime)\right],\label{tausame}\\
\frac{1}{\tau_\tau^{diff}(p_1)} &=& d\frac{(2\pi)^2}{(2\pi)^3} \int
p_2^2 dp_2 d\cos\theta_{12} d\cos\theta
\frac{d\sigma_{\tau,-\tau}}{d\Omega}\left|\frac{\vec{p}_1}{m_\tau^\star(p_1)}-\frac{\vec{p}_2}{m_{-\tau}^\star(p_2)}\right|
\notag\\
&\times&\left[n^0_{-\tau}(p_2)-n^0_{-\tau}(p_2)n^0_\tau(p_1^\prime)-n^0_{-\tau}(p_2)n^0_{-\tau}(p_2^\prime)+n^0_\tau(p_1^\prime)n^0_{-\tau}(p_2^\prime)\right],\label{taudiff}
\end{eqnarray}
\end{widetext}
where $\theta_{12}$ is the angel between $\vec{p}_1$ and
$\vec{p}_2$, and $\theta$ is the angel between the total momentum
$\vec{p}_{tot} = \vec{p}_1+\vec{p}_2 =
\vec{p}_1^\prime+\vec{p}_2^\prime$ and the relative momentum of the
final state $\vec{p}_{rel} = \vec{p}_1^\prime-\vec{p}_2^\prime$. As
for elastic collisions $|\vec{p}_1^\prime-\vec{p}_2^\prime|$ =
$|\vec{p}_1-\vec{p}_2|$, their magnitude can be calculated from
\begin{eqnarray}
|\vec{p}_{tot}| &=& \sqrt{p_1^2+p_2^2+2p_1p_2\cos\theta_{12}},\\
|\vec{p}_{rel}| &=& \sqrt{p_1^2+p_2^2-2p_1p_2\cos\theta_{12}}.
\end{eqnarray}
The magnitude of $\vec{p}_1^\prime$ and $\vec{p}_2^\prime$ can then
be obtained from
\begin{eqnarray}
|\vec{p}_1^\prime| &=& \frac{1}{2}\sqrt{p_{tot}^2+p_{rel}^2+2p_{tot}p_{rel}\cos\theta},\\
|\vec{p}_2^\prime| &=&
\frac{1}{2}\sqrt{p_{tot}^2+p_{rel}^2-2p_{tot}p_{rel}\cos\theta}.
\end{eqnarray}
We choose the isotropic nucleon-nucleon cross sections, and in free
space they are taken as the parameterized forms~\cite{Cha90}
\begin{eqnarray}
&&\sigma_{pp(nn)} = 13.73 - 15.04/v + 8.76/v^2 +
68.67v^4,\label{sigma1}\\
&&\sigma_{np} = -70.67 - 18.18/v + 25.26/v^2 +
113.85v,\label{sigma2}
\end{eqnarray}
where the cross sections are in mb and $v$ is the velocity of the
projectile nucleon with respect to the fixed target nucleon. The
above cross sections are the same as used in the IBUU transport
model calculations. Figure~\ref{freecs} shows the $pp$ and $np$
cross sections as functions of the center-of-mass energy of the two
colliding nucleons from the above parameterized form. This
parametrization describes very well the experimental data for the
beam energy from $10$ MeV to $1$ GeV~\cite{Cha90}, corresponding to
the center-of-mass energy $\sqrt{s}$ from $1883$ MeV to $2325$ MeV.
It is seen that for the most probable energies the cross section is
smaller for $pp$ collisions than for $np$ collisions.

As used in the IBUU transport model, by assuming all the matrix
elements of the nucleon-nucleon interaction are the same in free
space and in the medium~\cite{Pan92}, the in-medium nucleon-nucleon
cross sections are modified by the in-medium effective masses in the
following form~\cite{Li05}
\begin{equation}
\sigma^{medium}_{NN} =
\sigma_{NN}\left(\frac{\mu_{NN}^\star}{\mu_{NN}}\right)^2,
\end{equation}
where $\mu_{NN}$ ($\mu_{NN}^\star$) is the free-space (in-medium)
reduced mass of colliding nucleons.

Eqs.~(\ref{LHS}) and (\ref{RHS}) lead to
\begin{equation}\label{deltan}
\delta n_\tau(p) = \tau_\tau(p) \frac{\partial u_z}{\partial x}
\frac{p_z p_x}{p} \frac{d n^0_\tau}{dp}.
\end{equation}
As $F = -\eta \frac{\partial u_z}{\partial x}$, the shear viscosity
can be calculated from Eqs.~(\ref{shearF}) and (\ref{deltan}) as
\begin{eqnarray}\label{eta}
\eta &=& \sum_\tau -\frac{d}{(2\pi)^3} \int \tau_\tau(p)
\frac{p_z(p_z-mu_z)
p_x^2}{pm^\star_\tau} \frac{d n^0_\tau}{dp} dp_x dp_y dp_z \notag\\
&=& \sum_\tau -\frac{d}{(2\pi)^3} \int \tau_\tau(p) \frac{p_z^2
p_x^2}{pm^\star_\tau} \frac{d n^\star_\tau}{dp} dp_x dp_y dp_z
\end{eqnarray}
where $p=\sqrt{p_x^2+p_y^2+p_z^2}$ and $n^\star_\tau$ is the local
momentum distribution. The second equality sign comes from that the
shear viscosity is independent of the magnitude of the flow and
$u_z=0$ is chosen.

\section{Results and discussions}
\label{results}

\begin{figure}[tbh]
\includegraphics[scale=0.8]{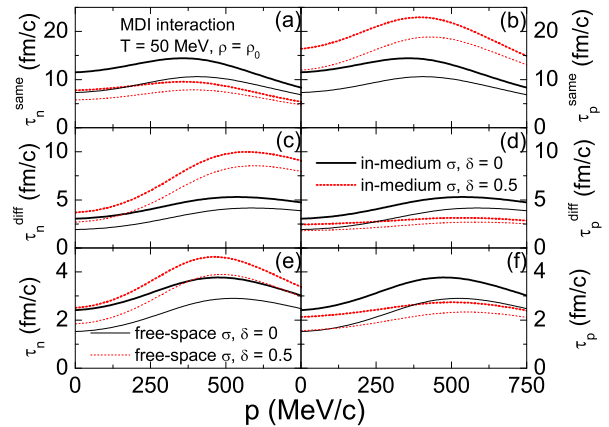}
\caption{(color online) Momentum dependence of the total relaxation
time and the relaxation time for the nucleon to collide with other
ones of same or different isospin in symmetric ($\delta=0$) and
neutron-rich ($\delta=0.5$) nuclear matter.} \label{tauisospin}
\end{figure}

The momentum dependence of the total relaxation time for a nucleon
and that for the nucleon to collide with other ones of same or
different isospin in symmetric and neutron-rich nuclear matter are
displayed in Fig.~\ref{tauisospin}. A smaller relaxation time means
the nucleon on average experiences more frequent collisions. A
constant cross section would make the relaxation time decrease with
increasing momentum, as higher-momentum nucleons are more likely to
collide with others. Using the energy-dependent free-space
nucleon-nucleon cross sections, there are peaks around $p=500$ MeV,
corresponding to the minimum values of the free cross sections as
shown in Fig.~\ref{freecs}. Due to the smaller effective masses of
nucleons in the nuclear medium, which leads to smaller in-medium
nucleon-nucleon cross sections, the relaxation times are larger. In
symmetric nuclear matter, $\tau_\tau$, $\tau^{same}_\tau$ and
$\tau^{diff}_\tau$ are the same for nucleons of different isospins.
In neutron-rich nuclear matter, $\tau_n^{same}$ are smaller while
$\tau_p^{same}$ are larger compared with that in symmetric nuclear
matter, due to larger chances for $nn$ collisions while smaller
chances for $pp$ collisions. In addition, $\tau_n^{diff}$ becomes
larger while $\tau_p^{diff}$ becomes smaller, due to a smaller
number of protons and a larger number of neutrons to collide with,
respectively. It is seen that there are always more chances for
collisions between nucleons with different isospins than those with
same isospin, mainly because of the larger degeneracy factor for
$\tau^{diff}$ than $\tau^{same}$ as in Eqs.~(\ref{tausame}) and
(\ref{taudiff}) and the larger cross section of
$\sigma_{\tau,-\tau}$ over $\sigma_{\tau,\tau}$. The total
relaxation time is dominated by the smaller one, i.e.,
$\tau^{diff}$. Thus $\tau_n>\tau_p$ is found in neutron-rich nuclear
matter, and they are both of the magnitude only a few fm/c at
$\rho=\rho_0$ and $T=50$ MeV.

\begin{figure}[tbh]
\includegraphics[scale=0.8]{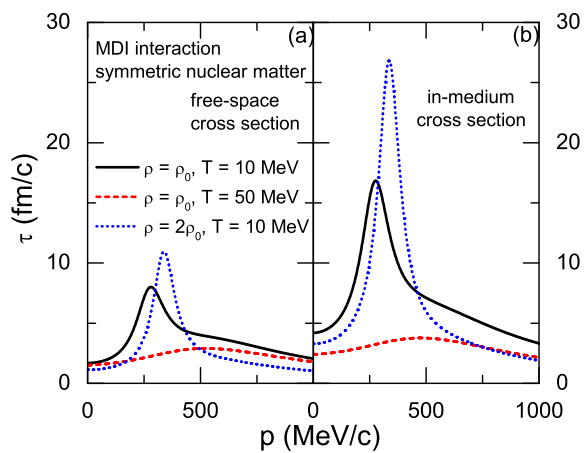}
\caption{(color online) Momentum dependence of the total relaxation
time at different densities and temperatures in symmetric nuclear
matter from free-space cross sections (left panel) and in-medium
cross sections (right panel).} \label{taurhoT}
\end{figure}

In Fig.~\ref{taurhoT} the effects of density and temperature on the
relaxation time are displayed. It is seen that compared with the
case at $T=50$ MeV, the relaxation time is generally larger at
$T=10$ MeV, due to the fact that the nucleons are less energetic at
lower temperatures. Furthermore, there appears a peak around $p=270$
MeV indicating a stronger Pauli blocking effect near the Fermi
surface. This can be understood as nucleons near the Fermi surface
are more likely to collide with those below the Fermi surface due to
larger cross sections at lower center-of-mass energy, but these
collisions are largely blocked at $T=10$ MeV. At $\rho=2\rho_0$ the
peaks move to a higher momentum and are even stronger, as the Fermi
momentum is higher and the Pauli blocking effect is stronger due to
the less diffusive distribution function as shown in Fig.~\ref{spp}.
Again the relaxation times are larger from smaller in-medium cross
sections, and at higher densities and lower temperatures this effect
is larger due to smaller nucleon effective masses.

\begin{figure}[tbh]
\includegraphics[scale=0.8]{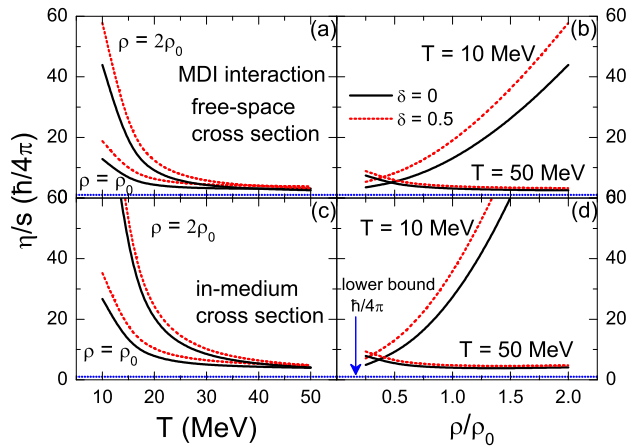}
\caption{(color online) Specific viscosity as a function of the
temperature at $\rho=\rho_0$ and $\rho=2\rho_0$ (left panels) and
that as a function of the density at $T=10$ MeV and $T=50$ MeV
(right panels) for symmetric ($\delta=0$) and neutron-rich
($\delta=0.5$) nuclear matter from free-space cross sections (upper
panels) and in-medium cross sections (lower panels). The lower bound
of the specific viscosity is also shown by dotted lines for
reference.} \label{etas}
\end{figure}

The specific viscosity, i.e, the ratio of the shear viscosity over
the entropy density, is shown in upper panels of Fig.~\ref{etas} as
a function of the temperature for different densities (panel (a))
and as a function of the density for different temperatures (panel
(b)) in symmetric and neutron-rich nuclear matter from free-space
nucleon-nucleon cross sections. The lower bound $\eta/s  \sim
\hbar/4\pi$ obtained by the Anti de Sitter/conformal field theory
(AdS/CFT) correspondence~\cite{Kov05} is also shown by dotted lines
for reference. The specific viscosity decreases with increasing
temperature, as a result of more frequent collisions and weaker
Pauli blocking effect at higher temperatures. Although the effective
mass decreases with increasing density, which leads to a larger flux
between flow layers as in Eq.~(\ref{shearF}), and a larger relative
velocity between nucleons in Eqs.~(\ref{tausame}) and
(\ref{taudiff}), the Pauli blocking effect is stronger at higher
densities. The combined effects lead to an increasing trend of the
specific viscosity with increasing density at lower temperatures,
but a slightly decreasing trend with increasing density at higher
temperatures when the Pauli blocking effect is much weaker.
Furthermore, it is interesting to seen that the specific viscosity
is larger in neutron-rich nuclear matter than in symmetric nuclear
matter at all the densities and temperatures, especially at lower
temperatures. This is because that the neutron relaxation time is
larger than proton in neutron-rich nuclear matter and it contributes
more to the total shear viscosity according to Eq.~(\ref{eta}) due
to a less diffusive neutron momentum distribution function. At lower
temperatures, this effect is larger because of the larger difference
between $\sigma_{\tau,\tau}$ and $\sigma_{\tau,-\tau}$, stronger
Pauli blocking effect on neutrons, and less diffusive momentum
distribution of neutrons as in Eq.~(\ref{eta}).

The density and temperature dependence of the specific viscosity
from in-medium cross sections are shown in the lower panels of
Fig.~\ref{etas}. Compared with the results from free-space
nucleon-nucleon cross sections, the specific viscosity is much
larger at lower temperatures and higher densities, when the
in-medium cross sections are smaller from smaller in-medium
effective masses. The isospin effect on the specific viscosity is
somehow smaller, compared with the results from free-space
nucleon-nucleon cross sections, due to the isospin-dependent
modification on the in-medium cross sections. As the effective mass
of neutrons is larger than that of protons in neutron-rich nuclear
matter, the isospin effect on $\tau_n$ is smaller from in-medium
cross sections than that from free-space cross sections, as shown in
panel (e) of Fig.~\ref{tauisospin}. At $T=50$ MeV, the specific
viscosity is only about $4\sim5$ $\frac{\hbar}{4\pi}$, which is
already close to the value of QGP extracted from the transverse
momentum spectrum and elliptic flow using the viscous hydrodynamic
model~\cite{Son11}. The specific viscosity obtained in this way is
similar to that extracted from the BUU model calculation using the
Green-Kubo formulas~\cite{Li11}, which is around $20 \sim 30
\frac{\hbar}{4\pi}$ at lower energies and reduces to as low as $6
\frac{\hbar}{4\pi}$ at higher energies. The density and temperature
dependence of the specific viscosity at lower temperatures are also
similar to those in Refs.~\cite{Dan84,Shi03},

\begin{figure}[tbh]
\includegraphics[scale=0.8]{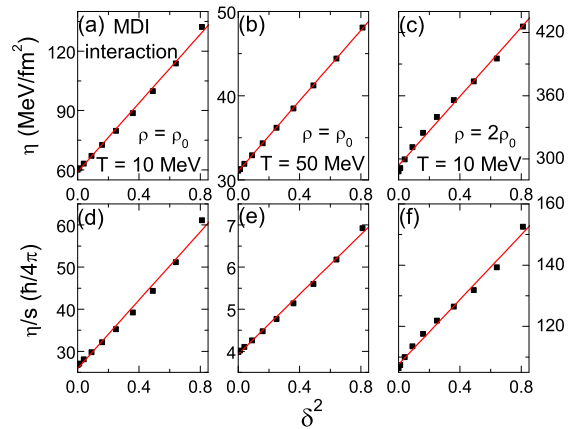}
\caption{(color online) Shear viscosity $\eta$ and specific
viscosity $\eta/s$ as functions of isospin asymmetry $\delta^2$ for
different densities and temperatures. The solid lines are from the
linear fit.} \label{parabolic}
\end{figure}

It will be interesting to study in detail how the shear viscosity
$\eta$ and the specific viscosity $\eta/s$ change with the isospin
asymmetry $\delta$, which is shown in Fig.~\ref{parabolic}. It is
seen that both $\eta$ and $\eta/s$ roughly satisfy the parabolic
approximation
\begin{eqnarray}
\eta(\rho,T,\delta) &\approx& \eta(\rho,T,\delta=0) +
\eta_{sym}(\rho_,T) \delta^2,\\
\left(\frac{\eta}{s}\right)(\rho,T,\delta) &\approx&
\left(\frac{\eta}{s}\right)(\rho,T,\delta=0) +
\left(\frac{\eta}{s}\right)_{sym}(\rho_,T) \delta^2,
\end{eqnarray}
where the coefficients $\eta_{sym}(\rho_,T)$ and
$\left(\frac{\eta}{s}\right)_{sym}(\rho_,T)$ are comparable to
$\eta(\rho,T,\delta=0)$ and
$\left(\frac{\eta}{s}\right)(\rho,T,\delta=0)$, respectively. The
large isospin effect is also observed in results from the Brueckner
theory~\cite{Zha10}. The parabolic approximation is good even for
very large isospin asymmetry, and it seems to be better at lower
densities or higher temperatures.

\section{Conclusions}
\label{summary}

In this paper I discussed the specific viscosity from the relaxation
time approach by using the MDI interaction and the nucleon-nucleon
cross sections from the experimental data modified by the in-medium
effective masses as used in the IBUU transport model calculations.
In neutron-rich nuclear matter, the relaxation time of neutrons
increases while that of protons decreases, compared with that in
symmetric nuclear matter. The specific viscosity decreases with
increasing temperature because of more frequent collisions and
weaker Pauli blocking effect. At lower temperatures, the specific
viscosity increases with increasing density due to increasing effect
of Pauli blocking while at higher temperatures it slightly decreases
with increasing density due the smaller in-medium effective masses.
The specific viscosity increases with increasing isospin asymmetry,
mainly from the larger relaxation time of neutrons in neutron-rich
nuclear matter. Both the shear viscosity and specific viscosity
roughly follow the parabolic approximation with respect to the
isospin asymmetry.

As in this frame work, the effect of the interaction on the specific
viscosity only comes from the effective mass, i.e., the momentum
dependence of the single-particle potential, the value of $x$, which
determines the density dependence of the symmetry energy, will not
affect the results. In reality, the transitive matrix
$T_{\tau,\tau^\prime}$ will also be modified in the medium,
especially at higher densities. In a more realistic
Brueckner-Hartree-Fock calculation~\cite{Zha07}, the in-medium cross
section is calculated by consistently taking account of the two-body
and three-body nucleon forces. It was found that the in-medium cross
section is more isotropic compared with the cross section in free
space as the forward and backward peaks are largely reduced, and
including the three-body force would further reduce the
nucleon-nucleon cross section. An isotropic cross section generally
leads to a smaller viscosity than an anisotropic one with forward
and backward peaks as the effective transport cross section is
larger in the former case, while including the contribution from the
three-body force seems to increase the viscosity. In addition, the
system is assumed to consist of uniform nuclear matter in the
present study. At higher densities and/or temperatures, inelastic
nucleon-nucleon collisions such as $NN \rightarrow N\Delta$ become
important, and $\Delta$ resonances and pions will be abundantly
produced. At lower densities and/or temperatures, liquid-gas phase
transition will occur and clusters will be formed. All these will
affect the viscosity of the system.

\begin{acknowledgments}
I thank Feng Li who is currently a graduate student in Cyclotron
Institute of Texas A\&M University for helpful discussions.
\end{acknowledgments}


\begin{thebibliography}{99}

\bibitem{Ber88} G.F. Bertsch and S. Das Gupta, Phys. Rep.
\textbf{160}, 189 (1988).

\bibitem{GBD87} C. Gale, G. Bertsch, and S. Das Gupta, Phys. Rev. C
\textbf{35}, 1666 (1987).

\bibitem{You99} D.H. Youngblood, H.L. Clark, and Y.W. Lui, Phys. Rev. Lett. \textbf{82}, 691 (1999).

\bibitem{Che05} L.W. Chen, C.M. Ko, and B.A. Li, Phys. Rev. Lett. \textbf{94}, 032701
(2005).

\bibitem{Yon06} G.C. Yong, B.A. Li, and L.W. Chen, Phys. Rev. C
\textbf{74}, 064617 (2006).

\bibitem{Yon09} G.C. Yong, B.A. Li, L.W. Chen, and X.C. Zhang, Phys.
Rev. C \textbf{80}, 044608 (2009).

\bibitem{Xia09} Z.G. Xiao, B.A. Li, L.W. Chen, G.C. Yong, and M. Zhang,
Phys. Rev. Lett. \textbf{102}, 062502 (2009).

\bibitem{Li08} B.A. Li, L.W. Chen, and C. M. Ko, Phys. Rep. \textbf{464}, 113 (2008).

\bibitem{Sch93} W. Schmidt, U. Katscher, B. Waldhauser, J.A. Maruhn,
H. St\"{o}cker, and W. Greiner, Phys. Rev. C \textbf{47}, 2782
(1993).

\bibitem{Pes05} A. Peshier and W. Cassing, Phys. Rev. Lett.
\textbf{94}, 172301 (2005).

\bibitem{Maj07} A. Majumder, B. M\"{u}ller, and X.N. Wang, Phys.
Rev. Lett. \textbf{99}, 192301 (2007).

\bibitem{Xu08} Z. Xu and C. Greiner, Phys. Rev. Lett. \textbf{100},
172301 (2008).

\bibitem{Che10} J.W. Chen, H. Dong, K. Ohnishi, and Q. Wang, Phys.
Lett. \textbf{B685}, 277 (2010).

\bibitem{Mur04} A. Muronga, Phys. Rev. C \textbf{69}, 044901 (2004).

\bibitem{Che07a} J.W. Chen and E. Nakano, Phys. Lett. \textbf{B647},
371 (2007).

\bibitem{Dem09} N. Demir and S.A. Bass, Phys. Rev. Lett.
\textbf{102}, 172302 (2009).

\bibitem{Pal10a} S. Pal, Phys. Lett. \textbf{B684}, 211 (2010).

\bibitem{Dan84} P. Danielewicz, Phys. Lett. \textbf{B146}, 168 (1984).

\bibitem{Shi03} L. Shi and P. Danielewicz, Phys. Rev. C \textbf{68},
064604 (2003).

\bibitem{Che07b} J.W. Chen, Y.H. Li, Y.F. Liu, and E. Nakano, Phys.
Rev. D \textbf{76}, 114011 (2007).

\bibitem{Pal10b} S. Pal, Phys. Rev. C \textbf{81}, 051601(R) (2010).

\bibitem{Li11} S.X. Li, D.Q. Fang, Y.G. Ma, and C.L. Zhou, Phys.
Rev. C \textbf{84}, 024607 (2011).

\bibitem{Zha10} H.F. Zhang, U. Lombardo, and W. Zuo, Phys. Rev. C
\textbf{82}, 015805 (2010).

\bibitem{Jeo95} S. Jeon, Phys. Rev. D \textbf{52}, 3591 (1995).

\bibitem{Kub66} R. Kubo, Rep. Prog. Phys. \textbf{29}, 255 (1966).

\bibitem{Hua87} K. Huang, {\it Statistical Mechanics}, 2nd edition (John Wiley \& Sons, New York, 1987).

\bibitem{Abu93} M.M. Abu-Samreh and H.S. K\"{o}hler, Nucl. Phys.
\textbf{A552}, 101 (1993).

\bibitem{Das03} C.B. Das, S. Das Gupta, C. Gale, and B.A. Li, Phys.
Rev. C \textbf{67}, 034611 (2003).

\bibitem{Xu10} J. Xu and C.M. Ko, Phys. Rev. C \textbf{82}, 044311 (2010).

\bibitem{Xu07} J. Xu, L.W. Chen, B.A. Li, and H.R. Ma, Phys. Rev. C \textbf{75}, 014607 (2007).

\bibitem{Li05} B.A. Li and L.W. Chen, Phys. Rev. C \textbf{72},
064611 (2005).

\bibitem{Li04} B.A. Li, Phys. Rev. C \textbf{69}, 064602 (2004).

\bibitem{Cha90} S.K. Charagi and S.K. Gupta, Phys. Rev. C \textbf{41}, 1610 (1990).

\bibitem{Pan92} V.R. Pandharipande and S.C. Pieper, Phys. Rev. C
\textbf{45}, 791 (1992).

\bibitem{Kov05} P.K. Kovtun, D.T. Son, and A.O. Starinets, Phys. Rev. Lett.
\textbf{94}, 111601 (2005).

\bibitem{Son11} H.C. Song, S.A. Bass, U. Heinz, T. Hirano, and C.
Shen, Phys. Rev. Lett. \textbf{106}, 192301 (2011).

\bibitem{Zha07} H.F. Zhang, Z.H. Li, U. Lombardo, P.Y. Luo, F.
Sammarruca, and W. Zuo, Phys. Rev. C \textbf{76}, 054001 (2007), and
references therein.

\end{thebibliography}
\end{document}